\documentclass[useAMS,usenatbib]{mn2e}
\usepackage{graphicx}
\usepackage[english]{babel}
\usepackage{amsmath}
\usepackage{amssymb}
\usepackage{textcomp}
\usepackage{natbib}
\title[The drop in the cosmic star formation rate below $z=2$]{The drop in the cosmic star formation rate below redshift~2 is caused by a change in the mode of gas accretion and by AGN feedback}
\author[F. van de Voort et al.]{Freeke van de Voort$^{1}$\thanks{E-mail: fvdvoort@strw.leidenuniv.nl},
Joop~Schaye$^1$,
C.~M.~Booth$^1$, and 
Claudio~Dalla~Vecchia$^{1,2}$\\
$^{1}$Leiden Observatory, Leiden University, Postbus 9513, 2300 RA, Leiden, The Netherlands\\
$^{2}$Max Planck Institute for Extraterrestrial Physics, Giessenbachstra\ss{}e 1, 85748 Garching, Germany
}

\begin{document}

\date{Accepted not yet. Received not yet; in original form \today}

\pagerange{\pageref{firstpage}--\pageref{lastpage}} \pubyear{2011}

\maketitle

\label{firstpage}

\begin{abstract}
The cosmic star formation rate is observed to drop sharply after redshift $z=2$. We use a large, cosmological, smoothed particle hydrodynamics simulation to investigate how this decline is related to the evolution of gas accretion and to outflows driven by active galactic nuclei (AGN). We find that the drop in the star formation rate follows a corresponding decline in the global cold-mode accretion rate density onto haloes, but with a delay of order the gas consumption time scale in the interstellar medium. Here we define cold-mode (hot-mode) accretion as gas that is accreted and whose temperature has never exceeded (did exceed) 10$^{5.5}$~K. In contrast to cold-mode accretion, which peaks at $z\approx 3$, the hot mode continues to increase to $z\approx 1$ and remains roughly constant thereafter. By the present time, the hot mode strongly dominates the global accretion rate onto haloes. Star formation does not track hot-mode halo accretion because most of the hot halo gas never accretes onto galaxies. AGN feedback plays a crucial role by preferentially preventing gas that entered haloes in the hot mode from accreting onto their central galaxies. Consequently, in the absence of AGN feedback, gas accreted in the hot mode would become the dominant source of fuel for star formation and the drop off in the cosmic star formation rate would be much less steep.  
\end{abstract}


\section{Introduction}

The cosmic star formation rate (SFR) is observed to peak between redshifts $z\approx3$ and $z\approx2$ after which it decreases by an order of magnitude \citep[e.g.][]{Hopkins2006}. Its evolution is thought to be determined by the combination of the formation and growth of (dark matter) haloes, which depends on cosmology, and the distribution of SFRs in haloes as a function of halo mass and redshift. The latter depends on processes such as gas accretion, stellar mass loss, radiative cooling, (re-)ionization, and feedback from star formation and active galactic nuclei (AGN) \citep[e.g.][]{Schaye2010}.

The observed rates of star formation in galaxies can only be sustained for long periods of time if the galaxies are being fed continuously \citep[e.g.][]{Bauermeister2010}. This feeding happens through the accretion of gas from the intergalactic medium. In the simplest picture of spherical collapse, it is assumed that all gas accreting onto a halo is shock-heated to the virial temperature of the halo, reaching a quasi-static equilibrium supported by the pressure of the hot gas. If the cooling time of the halo gas is sufficiently short, then it may subsequently enter the central galaxy. Whether or not
the halo gas can accrete onto the galaxy therefore depends on both the
temperature and the density of the gas \citep{Rees1977}. However, if the cooling time of gas that has gone through a (hypothetical) accretion shock at the virial radius is short compared with the age of the Universe, then there can be no hydrostatic halo and hence also no virial shock. This will be the case for low-mass haloes \citep{Rees1977, White1978}. The accretion rate onto the central galaxy then depends on the infall rate, but not on the cooling rate \citep{White1991}. Indeed, \citet{Birnboim2003} showed that a virial shock is absent for low-mass haloes in a spherically symmetric simulation.

However, cosmological simulations show significant deviations from spherical symmetry. Galaxies form inside the filaments that make up the cosmic web and these filaments contribute significantly to, or even dominate, the gas supply of galaxies \citep{Dekel2009a}. Because the density inside filaments is higher than that of the ambient gas, the cooling time of the gas is much shorter and the gas can accrete cold onto haloes of higher masses than predicted by spherically symmetric models \citep[e.g.][]{Keres2005, Dekel2006}. This cold gas can reach the high densities of the interstellar medium (ISM) much more efficiently than the tenuous, hot gas in the halo. It is this `cold mode' of accretion that predominantly feeds the galaxy and powers star formation \citep[][hereafter V10]{Keres2009a, Brooks2009, Voort2010}.

The fraction of the accreted gas that is accreted cold, i.e.\ the fraction of the gas that did not pass through a virial shock, depends on both halo mass and redshift (\citealt{Keres2005, Dekel2006, Ocvirk2008, Keres2009a, Brooks2009}; V10). In V10 we used simulations to show that the rate at which gas accretes onto central galaxies is generally much lower than the rate at which gas accretes onto their host haloes. Furthermore, we found that while halo accretion rates are determined by the depth of the gravitational potentials, galaxy accretion rates are also sensitive to processes such as metal-line cooling and feedback from star formation and AGN. 

Here we use a large cosmological hydrodynamical simulation that includes radiative cooling (computed element by element and thus including metal lines), star formation, stellar mass loss, and outflows driven by both supernovae and AGN, to determine how hot and cold accretion influence the global star formation history (SFH). We calculate global accretion rate densities, both for hot- and cold-mode accretion, and for accretion onto haloes as well as accretion onto galaxies. We will compare the evolution of these different global accretion rates to the resulting global SFH and learn how they all link together. We will show that the sharp drop in the global SFH after $z\approx2$ reflects the corresponding sharp drop in the rate of cold-mode accretion onto haloes. Motivated by semi-analytic models and simulations that have shown AGN feedback to be crucial for the suppression of star formation in high-mass haloes \citep[e.g.][]{Springeletal2005, Bower2006, Croton2006, Lagos2008, Booth2009, McCarthy2010}, we use an identical simulation, except for the omission of feedback from AGN, to investigate the effect of AGN feedback on the global accretion rates. We will show that the hot accretion mode is more strongly suppressed by AGN feedback than the cold mode. Without AGN feedback, low-redshift star formation would not be predominantly fuelled by the cold accretion mode and the drop in the cosmic SFR would be much less strong.

The outline of this paper is as follows. In Section~\ref{sec:sim} we briefly describe the simulations from which we derive our results and discuss our method for selecting recently accreted gas. In Section~\ref{sec:global} we compare global accretion rates to the cosmic SFR and show which haloes contribute most to the global accretion rates and the cosmic SFH. In Section~\ref{sec:REFAGN} we investigate the effect of AGN feedback on the hot and cold accretion rate densities. Finally, we summarize and discuss our results in Section~\ref{sec:concl}.

\section{Simulations} \label{sec:sim}

We use a modified version of \textsc{gadget-3} \citep[last described in][]{Springel2005}, a smoothed particle hydrodynamics (SPH) code that uses the entropy formulation of SPH \citep{Springel2002}, which conserves both energy and
entropy where appropriate. This work is part of the OverWhelmingly Large Simulations (OWLS) project \citep{Schaye2010}, which consists of a large number of cosmological simulations with varying (subgrid) physics. Here we make use of the so-called `AGN' and `reference' models, which are identical except that only model AGN includes supermassive black holes and AGN feedback. The AGN simulation will be our fiducial model, but we will compare it with the OWLS reference model in Section~\ref{sec:REFAGN}. As the simulations are fully described in \citet{Schaye2010}, we will only summarize their main properties here.

The simulations described here assume a $\Lambda$CDM cosmology with parameters $\Omega_\mathrm{m} = 1 - \Omega_\Lambda = 0.238$, $\Omega_\mathrm{b} = 0.0418$, $h = 0.73$, $\sigma_8 = 0.74$, $n = 0.951$. These values are consistent\footnote{The only significant discrepancy is in $\sigma_8$, which is 8 per cent, or 2.3$\sigma$, lower than the value favoured by the WMAP 7-year data.} with the WMAP year~7 data \citep{Komatsu2011}. 

A cubic volume with periodic boundary conditions is defined, within which the mass is distributed over $N^3$ dark matter and as many gas particles. The box size (i.e.\ the length of a side of the simulation volume) of the simulations used in this work is 100~$h^{-1}$ comoving Mpc, with $N=512$. The (initial) particle masses for baryons and dark matter are $1.2\times10^8$~M$_\odot$ and $5.6\times10^8$~M$_\odot$, respectively. 
The gravitational softening length is 7.8~$h^{-1}$comoving kpc, i.e.\ 1/25 of the mean dark matter particle separation, but we imposed a maximum of 2~$h^{-1}$proper kpc, which is reached at $z=2.91$.

The abundances of eleven elements released by massive stars and intermediate mass stars are followed as described in \citet{Wiersma2009b}.
We assume the stellar initial mass function (IMF) of \citet{Chabrier2003}, ranging from 0.1 to 100~M$_\odot$. As described in \citet{Wiersma2009a}, radiative cooling and heating are computed element by element in the presence of the cosmic microwave background radiation and the \citet{Haardt2001} model for the UV/X-ray background from galaxies and quasars.

Star formation is modelled according to the recipe of \citet{Schaye2008}. A polytropic equation of state $P_\mathrm{tot}\propto\rho_\mathrm{gas}^{4/3}$ is imposed for densities exceeding $n_\mathrm{H}=0.1$~cm$^{-3}$, where $P_\mathrm{tot}$ is the total pressure and $\rho_\mathrm{gas}$ the density of the gas. Gas particles with proper densities $n_\mathrm{H}\ge0.1$~cm$^{-3}$ and temperatures $T\le10^5$~K are moved onto this equation of state and can be converted into star particles. 
The star formation rate (SFR) per unit mass depends on the gas pressure and is set to reproduce the observed Kennicutt-Schmidt law \citep{Kennicutt1998}.

Feedback from star formation is implemented using the prescription of \citet{Vecchia2008}. About 40 per cent of the energy released by type II SNe is injected locally in kinetic form. The rest of the energy is assumed to be lost radiatively. The initial wind velocity is 600 km s$^{-1}$.

Our fiducial simulation includes AGN feedback. The model, which is a modified version of the model introduced by \citet{Springeletal2005}, is fully described and tested in \citet{Booth2009}. In short, a seed mass black hole is placed in every resolved halo. These black holes grow by accretion of gas, which results in the injection of energy in the surrounding medium, and by mergers.

The accretion rate onto the black hole equals the so-called Bondi-Hoyle accretion rate \citep{Bondi1944} if the gas density is low ($n_{\rm H} < 10^{-1}\,{\rm cm}^{-3}$). However, in dense, star-forming gas, where the accretion rate would be severely underestimated because the simulations do not include a cold, interstellar gas phase and because the Jeans scales are unresolved, the accretion rate is multiplied by an efficiency parameter, $\alpha$, given by $\alpha=(n_\mathrm{H}/n^*_\mathrm{H})^\beta$, where $n^*_\mathrm{H}=0.1$~cm$^{-3}$ is the threshold density for star formation and $\beta=2$.  Note, however, that our results are insensitive to the choice for $\beta$ as long as it is chosen to be large (see \citealt{Booth2009}). A fraction of 1.5 per cent of the rest-mass energy of the accreted gas is injected into the surrounding medium in the form of heat, by increasing the temperature of at least one neighbouring gas particle by at least $10^8$~K. The minimum temperature increase ensures that the feedback is effective, because the radiative cooling time of the heated gas is sufficiently long, and results in fast outflows. When AGN feedback is included, the SFR is reduced for haloes with $M_\mathrm{halo}\gtrsim10^{12}$~M$_\odot$ \citep{Booth2009}. The AGN simulation reproduces the observed mass density in black holes at $z=0$ and the black hole scaling relations \citep{Booth2009} and their evolution \citep{Booth2010} as well as the observed optical and X-ray properties, stellar mass fractions, SFRs, stellar age distributions and the thermodynamic profiles of groups of galaxies \citep{McCarthy2010}. 

The Lagrangian nature of the simulation is exploited by tracing each fluid element back in time, which is particularly convenient for this work, as it allows us to study the temperature history of accreted gas. During the simulations the maximum past temperature, $T_\mathrm{max}$, was stored in a separate variable. The variable was updated for each SPH particle at every time step for which the temperature exceeded the previous maximum temperature. The artificial temperature the particles obtain when they are on the equation of state (i.e.\ when they are part of the unresolved multiphase ISM) was ignored in this process. This may cause us to underestimate the maximum past temperature of gas that experienced an accretion shock at higher densities. Ignoring such shocks is, however, consistent with our aims, as we are interested in the maximum temperature reached \emph{before} the gas entered the galaxy.

Resolution tests are not included in this paper. However, \citet{Schaye2010} have shown that the box size and resolution of the reference simulation used in this paper suffice to obtain nearly converged results for the cosmic SFH at $z<3$, which changes by much less than a factor of 2 when changing the resolution by a factor of 8. At higher redshifts the global SFR density is, however, underestimated as a result of the finite resolution. Because AGN feedback moves the peak star formation activity to lower-mass haloes, the convergence of the AGN simulation may be slightly worse. V10 tested the convergence of accretion rates and the fraction of the accretion due to the hot mode as a function of halo mass. They found that increasing the resolution gives slightly higher cold accretion fractions, which would only strengthen the conclusions of this work. For quantities averaged over the entire simulation volume, we have to keep in mind that the minimum halo mass that is included depends on the resolution. The global fraction of gas accreted in the cold mode may therefore also depend on the resolution, because the cold fraction increases with decreasing halo mass. Increasing the resolution would allow us to include lower-mass haloes, thus increasing the global cold accretion fraction. Again, this would only strengthen our conclusions.

\subsection{Accretion and mergers} \label{sec:acc}

In this section we summarize our method for determining the gas accretion rates onto haloes and galaxies and the amount of star formation within haloes. Our method is described in more detail in V10. We use \textsc{subfind} \citep{Dolag2009} to identify gravitationally bound haloes and subhaloes. In this work we only investigate accretion onto haloes that are not embedded inside larger haloes. In other words, we do not consider subhaloes. The simulations are saved at discrete output redshifts, called snapshots. The redshift intervals between snapshots are $\Delta z=0.25$ for $0< z\le 4$, and $\Delta z=0.5$ for $4< z\le 6$. This is the time resolution for determining accretion rates. For all haloes, called descendants, with more than 100 dark matter particles (i.e.\ with $M_\mathrm{halo}\gtrsim10^{10.8}$~M$_\odot$), we find the progenitor at the previous output redshift. We do this by finding the group containing most of the 25 most bound dark matter particles of the descendant. Haloes without a well-defined progenitor are discarded from our analysis. All gas entering a halo, both from accretion and from mergers, contributes to its growth. We therefore identify all the gas and recently formed star particles that are in the descendant, but not in its progenitor. 

To participate in star formation, gas has to enter not only the halo, but also the ISM of the galaxy. For the same haloes as we have selected above, we identify all particles entering the ISM (i.e.\ moving onto the equation of state) of the descendant galaxy between the two consecutive output redshifts. Galaxy mergers are therefore automatically included. To make sure we are not looking at accretion onto substructures within the main halo, we only consider accretion taking place within the inner 15 per cent of the virial radius. This corresponds to the growth of the central galaxy.

Finally, we identify all the stars formed in the same haloes and redshift bins to calculate the corresponding SFRs. Again, unresolved haloes and satellites are excluded.

We decided to include gas contributed by mergers, as well as smoother accretion flows, because both are mechanisms by which haloes and galaxies grow and both can provide fuel for star formation. We note, however, that the contribution from mergers is much smaller than that from smooth accretion. Mergers with mass ratios greater than 1:10 account for only $\la 10$ per cent of the total halo growth, except in groups and clusters at low redshift (see V10).

The simulations used for this study do not resolve haloes below $M_\mathrm{halo}=10^{10.8}$~M$_\odot$. We are therefore missing contributions to the global accretion rates and SFR from lower mass haloes. Because we use a fixed maximum past temperature threshold of $10^{5.5}$~K to distinguish hot from cold accretion (see below), small haloes with virial temperatures below this threshold will by definition accrete almost all gas in the cold mode. The mass regime that we investigate here is therefore the most interesting regime, where changes in accretion mode are expected to occur \citep{Dekel2006}.

\section{Global accretion and star formation} \label{sec:global}

\citet{Schaye2010} showed that without feedback from supernovae and AGN, the SFR density is overpredicted by a large factor at $z\lesssim2$. By including supernova feedback they could lower the SFR density, but if the predicted SFR density matched the observed peak at $z\approx2$, then it overpredicted the SFR density at $z=0$. The drop in the global SFR below $z=2$ is much closer to the observed slope if AGN feedback is included, but in that case the SFR density may be slightly too low \citep{Schaye2010}. This discrepancy could be solved by decreasing the efficiency of feedback from star formation, which had been set to reproduce the observed peak in the SFH using models without AGN feedback. It is, however, not clear how seriously we should take the discrepancy, since the AGN simulation does reproduce the observed masses and ages of the central galaxies of groups \citep{McCarthy2010}. This would suggest that the problem may be solved by increasing $\sigma_8$ from 0.74 to the currently favoured value of 0.81, which would have a relatively strong effect on global star formation rates \citep[see][]{Schaye2010} while leaving the evolution of haloes of a given mass nearly unchanged. Because, for the purposes of this work, the AGN simulation is the most realistic run from the OWLS suite, we will adopt it as our fiducial model. In Section~\ref{sec:REFAGN} we will discuss how its predictions differ from those of a simulation without AGN feedback.

\begin{figure*}
\center
\includegraphics[scale=0.58]{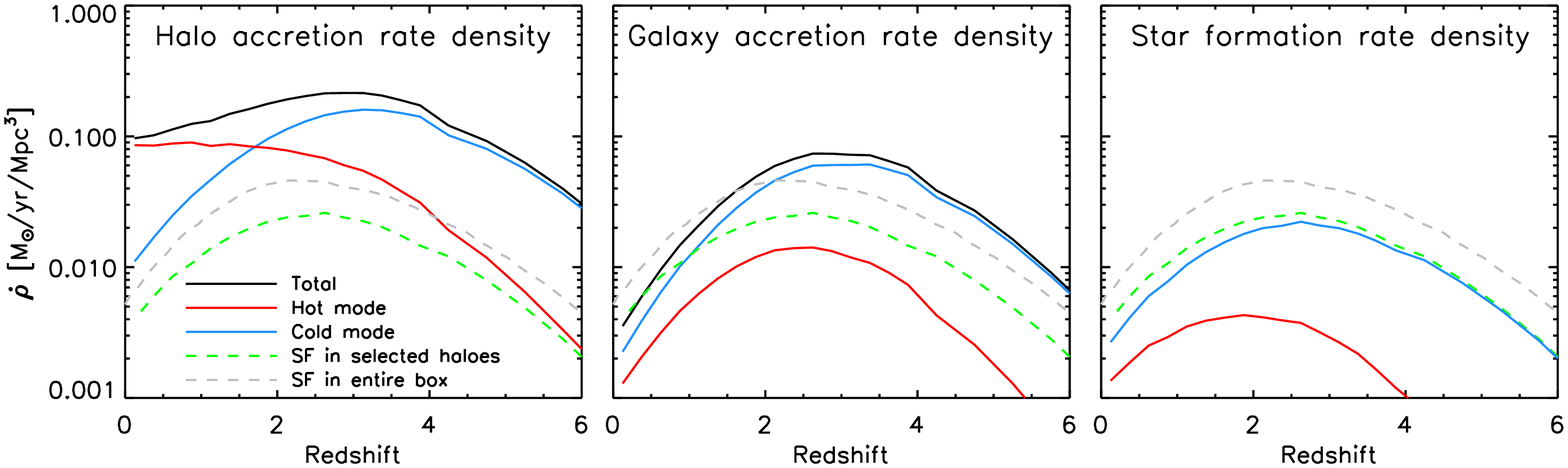}
\caption {\label{fig:global} Evolution of global accretion rate and SFR densities in resolved haloes with well-defined progenitors.
The left, middle, and right panels show global accretion rate densities onto haloes, onto central galaxies, and global SFR densities, respectively. Red and blue curves show accretion rate densities (left and middle panels) and SFR densities (right panel) resulting from hot- and cold-mode accretion, respectively. In all panels, the black curve is the sum of the red and blue curves, the green curve shows the global SFR density in the selected haloes, and the grey curve shows the global SFR density in the entire simulation box. The small `step' visible at $z\approx4$ is caused by the sudden increase in the time resolution for determining accretion, i.e.\ $\Delta z$ between snapshots decreases by a factor of two at $z=4$. Galaxies accrete most of their gas in the cold
mode and this mode is responsible for an even larger fraction of the star formation. Because of outflows driven by supernovae and AGN, the SFR density is generally lower than the galaxy accretion rate density. The global SFR declines more rapidly than either the total and hot-mode accretion rate densities. This decline must therefore be caused by the drop in global cold-mode accretion rate, though with a delay.}
\end{figure*}
\begin{figure*}
\center
\includegraphics[scale=0.58]{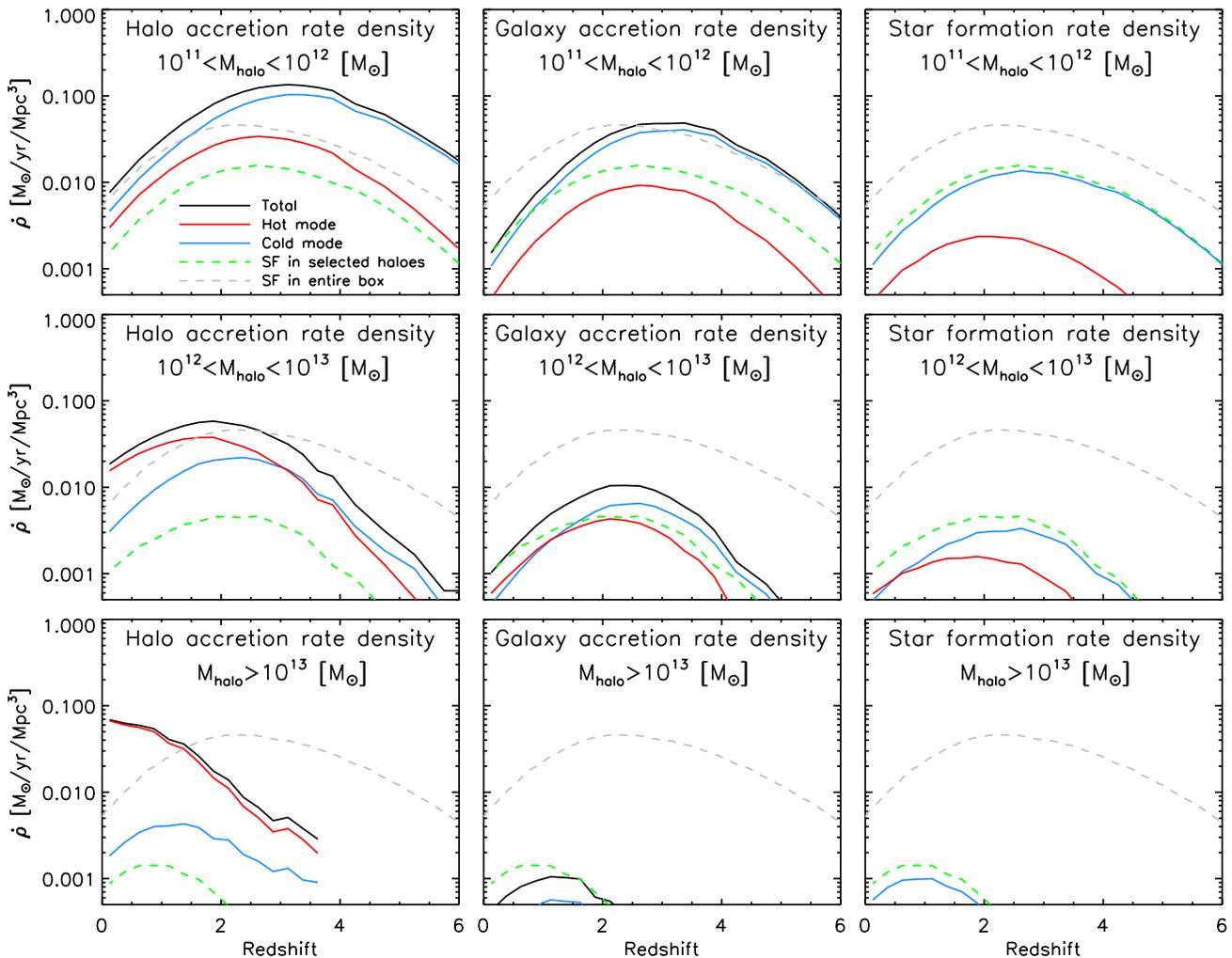}
\caption {\label{fig:globalmass} Evolution of global accretion rate densities onto haloes (left column), galaxies (middle column), and global SFR densities (right column) for different halo mass bins. From top to bottom only haloes have been included with: $10^{11}\le M_\mathrm{halo}<10^{12}$~M$_\odot$, $10^{12}\le M_\mathrm{halo}<10^{13}$~M$_\odot$, and $M_\mathrm{halo}\ge10^{13}$~M$_\odot$. The curves show the same quantities as in Figure~\ref{fig:global}. Above $z\approx3.5$ the highest mass bin contains no haloes. At all redshifts most of the cold halo accretion, galaxy accretion, and star formation happens in low-mass haloes (i.e.\ $M_\mathrm{halo}<10^{12}$~M$_\odot$). At $z\gtrsim2$ low-mass haloes also dominate the global hot halo accretion rates. At $z\lesssim1$ the total halo accretion rate is dominated by high-mass haloes, but nearly all of the gas is accreted hot and unable to accrete onto galaxies.}
\end{figure*}
\begin{figure*}
\center
\includegraphics[scale=0.58]{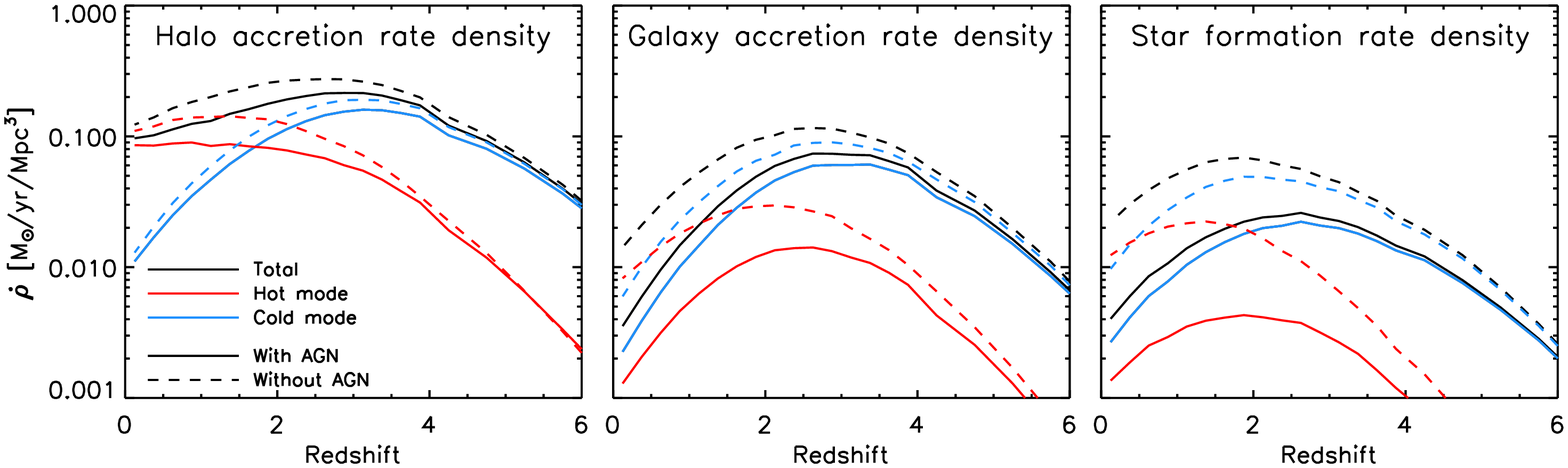}
\caption {\label{fig:globalREFAGN} Evolution of global accretion rate densities onto haloes (left column), galaxies (middle column), and global SFR densities for simulations with (solid curves, same as in Figure \ref{fig:global}) and without (dashed curves) AGN feedback. The black, red, and blue curves show the global accretion and SFR densities from all, hot-mode, and cold-mode accretion, respectively. AGN feedback suppresses halo accretion only slightly, but the effect on galaxy accretion and star formation is large, up to an order of magnitude. AGN feedback preferentially suppresses \textsl{hot}-mode galaxy accretion and star formation from gas accreted in the \textsl{hot} mode.}
\end{figure*}

The temperature of accreting gas has been found to follow a bimodal distribution (e.g.\ \citealt{Keres2005}; V10). Therefore, two modes of accretion are considered: the hot and the cold mode. The cooling function peaks at about $T=10^{5-5.5}$~K \citep[e.g.][]{Wiersma2009a}, so there is little gas around this temperature. We therefore define hot (cold) accretion as accreted gas with maximum past temperatures above (below) $10^{5.5}$~K. Much of the gas that is shock-heated to much higher temperatures is expected to stay hot for a long time, whilst gas that is heated to lower temperatures can more easily cool and condense onto the galaxy.

When considering accretion onto galaxies, it is important to note that the terms `hot' and `cold' refer to the maximum past temperature rather than the current temperature. In fact, gas that has been shock-heated to temperatures in excess of $10^{5.5}$~K \emph{must} cool down before it can accrete onto a galaxy, as we identify this type of accretion as gas joining the ISM, which in our simulations means that the gas density crosses the threshold of $n_\mathrm{H}=0.1$~cm$^{-3}$ while the temperature $T\le10^5$~K. Thus, although gas that, according to our terminology (which is consistent with that used in the literature), is accreted hot onto galaxies has been hot in the past, it is cold at the time of accretion. In V10 we showed that the
maximum past temperature is usually reached around the time the gas accreted onto the halo. 

The global accretion rate density onto \textit{haloes}, i.e.\ the gas mass accreting onto resolved haloes per year and per comoving Mpc$^3$, is shown in the left panel of Figure~\ref{fig:global} by the solid, black curve. The solid, red and blue curves show global accretion rates for hot- and cold-mode accretion, respectively. These global accretion rate densities are averaged over the time interval between two snapshots. The SFR density in the haloes we consider is shown by the green curve. The SFH in the entire box is shown by the dashed, grey curve. It is higher than the dashed, green curve, because the latter excludes star formation in subhaloes and unresolved haloes (i.e.\ $M_\mathrm{halo}<10^{10.8}$~M$_\odot$).

The global accretion rate onto resolved main haloes (solid, black curve) peaks at $z\approx3$. It is fairly constant, varying only by about a factor of two from $z\approx4$ down to $z\approx0$. The average accretion rate onto haloes of a given mass decreases more strongly towards lower redshift (V10). However, the number of haloes at a fixed mass increases and higher mass haloes form with decreasing redshift. The combination of these effects results in an almost constant global accretion rate density. We note, however, that the normalization and shape are not fully converged with respect to resolution: higher resolution simulations may find higher total and cold accretion rate densities. Increasing the resolution would allow us to include haloes with lower masses, which would boost the cold accretion rate density by up to a factor of $\sim2-3$. The total accretion rate density would therefore increase at $z\gtrsim2$, where cold accretion dominates. 
The global accretion rate is an order of magnitude higher than the global SFR in the same haloes (green, dashed curve), indicating that most of the gas that accretes onto haloes never forms stars. 

The global growth of haloes is dominated by cold accretion for $z>2$, but by $z=0$ the contribution of the hot mode exceeds that of the cold mode by an order of magnitude. The cold accretion rate density peaks around $z=3$ and falls rapidly thereafter, while the hot accretion rate density increases down to $z\approx2$ and flattens off at lower redshifts. The global SFR peaks around $z=2$ and declines by an order of magnitude to $z=0$. An order of magnitude drop is also visible for the global cold accretion rate from $z=3$ to $z=0$. Neither the total nor the hot accretion rate histories can explain the drop in the cosmic SFH, which must therefore be driven by the drop in the cold accretion rate.

The middle panel of Figure~\ref{fig:global} shows the global accretion rates onto central \textit{galaxies}. The green and grey curves are identical to the ones shown in the left panel. The black, red, and blue curves describe the total, hot, and cold accretion rate densities, respectively. The total accretion rate shows the amount of gas that joins the ISM. The gas can, however, be removed from the ISM by supernova and AGN feedback, as well as by dynamical processes. This is why the overall normalization of the SFR is generally lower than the ISM accretion rate. The global SFR peaks later than the global galaxy accretion rate ($z\approx 2$ versus $z\approx 3$, which corresponds to a difference of $\sim1$~Gyr in the age of the Universe). This delay probably results from the time it takes to convert interstellar gas into stars. The long gas consumption time scale implied by the assumed, Kennicutt star formation law ($\sim1$~Gyr for typical densities) allows the existence of reservoirs of accreted gas. The SFR can therefore temporarily be higher than the galaxy accretion rate, as happens for $z<0.5$ (compare the dashed, green curve with the solid, black curve, which include the same sample of resolved haloes). Gas returned to the ISM by stellar mass loss, a process that is included in our simulations and that becomes important for $z<1$ \citep{Schaye2010, Leitner2010}, also increases the SFR relative to the accretion rate.

As we did in the left panel of Figure~\ref{fig:global} for accretion onto haloes, we split the global accretion rate onto galaxies into separate contributions from the hot and cold modes. The global hot accretion rate peaks around $z=2$, as does the SFR density. The hot accretion rate is, however, not nearly enough to account for all the star formation in these galaxies, falling short by at least a factor of 2 at all redshifts. At $z>3$, the global cold accretion rate is an order of magnitude higher than the global hot accretion rate. This difference decreases to $\sim0.25$~dex by $z=0$. At all redshifts, it is mostly cold accretion that allows for the growth of \textit{galaxies}, even though hot accretion dominates the growth of \textit{haloes} below $z\approx2$.

Comparing the middle panel to the left panel, we notice that the global cold accretion rate onto galaxies is a factor of $\sim3-4$ lower than the cold accretion rate onto haloes. Not all cold gas that accretes onto haloes makes it into the central galaxy to form stars (see also V10). The shapes of the blue curves are, however, similar, indicating that the fraction of the gas accreting cold onto the halo that proceeds to accrete onto the galaxy is roughly constant with time. 

The situation is very different for global hot accretion rates (red curves). Hot accretion onto the ISM has already peaked at $z\approx2.5$, while hot accretion onto haloes continues to increase down to $z=0$.
This can be explained by noting that as the Universe evolves, gas is heated to higher temperatures, because haloes become more massive. In addition, the average density of the Universe goes down. The lower densities and higher temperatures gives rise to longer cooling times.
Moreover, winds from supernovae and AGN eject low-entropy gas at high redshift, raising the entropy of the gas in haloes at low redshift \citep{Crain2010, McCarthy2011}. Hence, as time goes on, more of the halo gas is unable to cool and reach the central galaxy. While the gravitational potential is the most important factor for the growth of haloes, for the growth of galaxies, the cooling function and feedback processes also come into play.

The right panel of Figure \ref{fig:global}, which shows the global SFR densities due to hot and cold accretion, confirms that the main fuel for star formation is gas accreted in the cold mode. The difference becomes smaller towards lower redshift. However, even at $z=0$ hot mode gas contributes 0.3~dex less than cold mode gas.

To investigate which haloes contribute most to the global accretion rates and SFHs, we show the same quantities as in Figure~\ref{fig:global} for three different halo mass bins in Figure~\ref{fig:globalmass}. From top to bottom, the mass ranges are $10^{11}\le M_\mathrm{halo}<10^{12}$~M$_\odot$, $10^{12}\le M_\mathrm{halo}<10^{13}$~M$_\odot$, and $M_\mathrm{halo}\ge10^{13}$~M$_\odot$, which contain 21813, 2804, and 285~haloes at $z=0$. The shape of the total halo accretion rate density in the lowest mass range is in agreement with that found by \citet{Bouche2010} based on an extended Press-Schechter formalism and a fit to dark matter accretion rates in N-body simulations.

For $z>2$ the global accretion rate densities onto haloes (left column), galaxies (middle column), and the global star formation rate density (right column) are all dominated by haloes with $M_\mathrm{halo}<10^{12}$~$M_\odot$ (top row). At that time, higher-mass haloes are still too rare to contribute significantly. Below $z\approx2$ haloes with $M_\mathrm{halo}=10^{12-13}$~M$_\odot$ (middle row) begin to contribute significantly and for accretion onto haloes, but not for accretion onto galaxies or star formation, their contribution is overtaken by $M_\mathrm{halo}=10^{13-14}$~M$_\odot$ haloes (bottom row) around $z=1$. 

Observe that the global halo accretion rate density starts to decline at $z\approx3$ and $z\approx2$ for the low and middle halo mass bins, respectively, and that it keeps increasing down to $z=0$ for $M_\mathrm{halo}\ge10^{13}$~M$_\odot$. The global cold accretion rate density decreases with time for $z<3$, $z<2.5$, and $z<1$ for the low, middle, and high halo mass bins, respectively.

Both the galaxy accretion rate density and the SFR density are dominated by $M_\mathrm{halo}<10^{12}$~M$_\odot$ haloes at all redshifts. Towards lower redshifts, high-mass haloes account for larger fractions of the total galaxy accretion rate density and SFR density, though they never dominate. High-mass haloes do dominate the halo accretion rate at low redshift, but nearly all of the gas is accreted hot and only a very small fraction of this gas is subsequently able to cool down onto galaxies.

\section{Effect of AGN feedback} \label{sec:REFAGN}

It is interesting to see what the influence of AGN feedback is on our results. Because AGN feedback is more important in higher-mass haloes, for which hot-mode accretion is more important, we expect it to have a larger effect on the hot-mode accretion rate density. Moreover,
it has been hypothesized \citep[e.g.][]{Keres2005, Dekel2006} that hot, dilute gas may be more vulnerable to AGN feedback than cold streams and may therefore be preferentially prevented from accreting onto galaxies. Indeed, \citet{Theuns2002} had already demonstrated that supernova-driven outflows follow the path of least resistance, leaving the cold filaments that produce HI absorption intact.

\citet{McCarthy2011} have shown that feedback from AGN at high redshift increases the entropy of the halo gas at low redshift. The hot gas will therefore be even hotter and less dense at low redshift than it would be in the absence of AGN feedback, making it more susceptible to being heated or entrained in an outflow, and thus to being prevented from accreting.

We compare our fiducial simulation, which includes AGN feedback, to the OWLS `reference model' which is identical to our fiducial run except that it does not include black holes and AGN feedback. This allows us to assess the effect of AGN feedback on the global hot and cold accretion rates. Figure~\ref{fig:globalREFAGN} shows the same solid curves as were shown in Figure~\ref{fig:global}. They indicate the total, hot, and cold accretion rate densities onto haloes (left panel) and onto galaxies (middle panel) and the star formation rate density resulting from all, hot, and cold accretion (right panel). The dashed curves show the same global accretion rates and SFHs for the simulation without AGN feedback. For accretion onto galaxies and for star formation the differences are striking. When AGN feedback is excluded, late-time star formation is no longer predominantly fuelled by gas accreted in the cold mode.

As expected, all accretion rate densities are reduced by the inclusion of AGN feedback. The effect on halo accretion is, however, small, as was also shown by V10. The hot and cold halo accretion rate densities are reduced by at most 0.2~and 0.1~dex, respectively. This reduction implies that AGN feedback also affects some gas outside of haloes. Even though the effect is small, AGN feedback reduces hot halo accretion more than cold halo accretion. 

The differential effect of AGN feedback on hot and cold accretion is much more pronounced for accretion onto galaxies than for accretion onto haloes and it increases towards lower redshift. At very high redshift ($z=9$), \citet{Powell2010} have shown that outflows (driven by supernova feedback) do not affect the galaxy inflow rates. Our results indicate that this may change towards lower redshifts, when densities are much lower. While AGN feedback reduces cold accretion rate densities by up to 0.4~dex (at $z=0$), the hot accretion rate densities decrease by up to 0.8~dex (also at $z=0$). The SFR densities are reduced by up to 0.6~dex for star formation powered by cold-mode accretion, but by 1~dex for hot-mode accretion. The reduction due to AGN feedback is thus $\sim0.4$~dex greater for the hot mode than for the cold mode, both for galaxy accretion and for star formation. The larger reduction indicates that AGN feedback preferentially, but not exclusively, prevents hot mode gas from accreting onto galaxies and participating in star formation. 

Hence, the inclusion of AGN feedback strongly boosts the size of the drop in the cosmic SFR at late times.
This preferential suppression of hot accretion is the result of two effects, namely of the differential effect at a fixed halo mass, indicating that hot-mode gas is more vulnerable to feedback than cold-mode gas, and of the fact that AGN feedback is effective only in massive haloes (with $M_\mathrm{halo}\gtrsim10^{12}$), for which hot accretion is important. The latter is the dominant effect.

\section{Conclusions} \label{sec:concl}

We have investigated the evolution of the global gas accretion rate densities onto haloes and onto their central galaxies and we have done so for both the hot and cold accretion modes. In addition, we studied the contributions from gas accreted through the cold and hot modes to the cosmic star formation history. We made use of a 100 Mpc/h, $2\times 512^3$ particle SPH simulation from the OWLS project that includes radiative cooling (computed element by element and thus including metal lines), star formation, stellar mass loss, supernova feedback, and AGN feedback. We isolated the effect of AGN feedback by comparing to a second simulation that did not include AGN, but which was otherwise identical. The hot and cold accretion modes were separated by using a fixed maximum past temperature threshold of $T_\mathrm{max}=10^{5.5}$~K.

The global gas accretion rate density onto haloes is much higher than that onto galaxies and both rates exceed the cosmic SFR density. This confirms the finding of V10 that most of the gas accreting onto haloes does not result in star formation. This is the case for both accretion modes, but the differences are larger for the hot mode. 

The global SFR declines after $z\approx2$, whereas the global hot-mode accretion rate onto haloes shows no such trend. From this, we conclude that the global SFR follows the drop in the global cold-mode accretion rate onto haloes, which sets in at $z\approx3$, but with a delay of order the gas consumption time scale in the ISM. Star formation tracks cold-mode accretion rather than hot-mode accretion because cold streams can reach the central galaxy, where star formation takes place, much more easily than gas that is shock-heated to high temperatures near the virial radius. Much of the hot gas cannot cool within a Hubble time and therefore cannot accrete onto the central galaxy. In addition, we demonstrated that it is very important that hot gas is more susceptible to removal by outflows driven by feedback from AGN. Without AGN feedback, gas accreted in the hot mode contributes significantly to the cosmic SFR below $z=1$ and the drop in the SFR below $z=2$ would be much smaller.

For the hot mode the difference between the accretion rates onto haloes and onto galaxies is larger at lower redshifts. While the hot accretion mode dominates the growth of haloes by an order of magnitude at $z\approx0$, it is still less important than cold accretion for the growth of the central galaxies. At $z > 2$, cold accretion even dominates the global accretion rate onto haloes. 

We demonstrated that AGN feedback suppresses accretion onto galaxies
and that it does so much more efficiently for the hot mode than for the cold mode. This happens because AGN feedback only becomes more efficient than feedback from star formation in high-mass haloes, which are dominated by hot accretion, and because hot-mode gas is more dilute and therefore more vulnerable to feedback. In addition, as demonstrated by 
\citet{McCarthy2011}, by ejecting low-entropy halo gas at high redshift ($z \ga 2$), AGN feedback results in an increase of the entropy, and thus a reduction of the cooling rates, of hot halo gas at low redshift. 

While \citet{Keres2009a} did not investigate accretion onto haloes, they did also find that cold accretion is most important for the growth of galaxies, with hot accretion becoming increasingly important towards lower redshifts (see also \citealt{Keres2005,Ocvirk2008,Brooks2009}; V10). However, their simulation included neither winds from supernovae nor feedback from AGN. AGN feedback was in fact ignored by all previous cosmological simulations investigating gas accretion except for \citet{Khalatyan2008}, who simulated a single object, and except for V10. Our results suggest that the neglect of this important process leads to a strong overestimate of the global accretion rate and SFR densities and of the importance of the hot accretion mode for galaxy accretion and star formation.

In summary, the rapid decline in the cosmic SFR density below $z=2$ is driven by the corresponding drop in the cold accretion rate density onto haloes. The total accretion rate onto haloes falls off much less rapidly because the hot mode becomes increasingly important. AGN feedback, which acts preferentially on gas accreted in the hot mode, prevents the hot halo gas from accreting onto galaxies and forming stars and is therefore a crucial factor in the steep decline of the cosmic SFR density.

\section*{Acknowledgements}

We would like to thank Avishai Dekel and all the members of the OWLS team for valuable discussions and the anonymous referee for useful comments. The simulations presented here were run on Stella, the LOFAR BlueGene/L system in Groningen, on the Cosmology Machine at the Institute for Computational Cosmology in Durham as part of the Virgo Consortium research programme, and on Darwin in Cambridge. This work was sponsored by the National Computing Facilities Foundation (NCF) for the use of supercomputer facilities, with financial support from the Netherlands Organization for Scientific Research (NWO), also through a VIDI grant, and from the Marie Curie Initial Training Network CosmoComp (PITN-GA-2009-238356).

\bibliographystyle{mn2e}
\bibliography{globalAGN}

\bsp

\label{lastpage}

\end{document}